\documentclass[conference]{IEEEtran}

\usepackage{scalerel}
\usepackage{tikz}
\usetikzlibrary{svg.path}

\definecolor{orcidlogocol}{HTML}{A6CE39}
\tikzset{
    orcidlogo/.pic={
        \fill[orcidlogocol] svg{M256,128c0,70.7-57.3,128-128,128C57.3,256,0,198.7,0,128C0,57.3,57.3,0,128,0C198.7,0,256,57.3,256,128z};
        \fill[white] svg{M86.3,186.2H70.9V79.1h15.4v48.4V186.2z}
        svg{M108.9,79.1h41.6c39.6,0,57,28.3,57,53.6c0,27.5-21.5,53.6-56.8,53.6h-41.8V79.1z M124.3,172.4h24.5c34.9,0,42.9-26.5,42.9-39.7c0-21.5-13.7-39.7-43.7-39.7h-23.7V172.4z}
        svg{M88.7,56.8c0,5.5-4.5,10.1-10.1,10.1c-5.6,0-10.1-4.6-10.1-10.1c0-5.6,4.5-10.1,10.1-10.1C84.2,46.7,88.7,51.3,88.7,56.8z};
    }
}

\usepackage{amsmath,amssymb,amsfonts}
\usepackage{algorithm}
\usepackage{algpseudocode}
\usepackage{booktabs}

\usepackage{subcaption}
\usepackage{enumitem}
\usepackage{wrapfig}
\usepackage{textcomp}
\usepackage{xcolor}
\usepackage{caption}
\usepackage{tabularx}
\usepackage{multirow}
\newcommand{\tp}{\mathsf{T}}

\def\BibTeX{{\rm B\kern-.05em{\sc i\kern-.025em b}\kern-.08em
    T\kern-.1667em\lower.7ex\hbox{E}\kern-.125emX}}

\newcommand\orcidicon[1]
{
    \href{https://orcid.org/#1}{\mbox{\scalerel*{
                \begin{tikzpicture}[yscale=-1,transform shape]
                \pic{orcidlogo};
                \end{tikzpicture}
            }{|}}}}
\usepackage{hyperref}

\begin{document}

\title{Revealing and Utilizing In-group Favoritism for Graph-based Collaborative Filtering}

\author{
    \IEEEauthorblockN{
        Hoin Jung\textsuperscript{1*}, 
        Hyunsoo Cho\textsuperscript{2*}, 
        Myungje Choi\textsuperscript{1*}, 
        Joowon Lee\textsuperscript{2*}, 
        Jung Ho Park\textsuperscript{1*}, 
        Myungjoo Kang$\text{\orcidicon{0000-0002-8064-7167}}$\textsuperscript{2$\dagger$}
    }
    \IEEEauthorblockA{
        \textsuperscript{1}\textit{Department of Computational Science and Engineering, Seoul National University}\\
        \textsuperscript{2}\textit{Department of Mathematical Sciences, Seoul National University}\\
        Seoul, Republic of Korea \\
        \{hoyin, hscho100, dnmslyyd, powep, jhpark009, mkang\}@snu.ac.kr
    }
}

\maketitle
\def\thefootnote{*}\footnotetext{These authors contributed equally to this work}\def\thefootnote{\arabic{footnote}}
\def\thefootnote{$\dagger$}\footnotetext{\textit{Member, IEEE}}\def\thefootnote{\arabic{footnote}}

\begin{abstract}
When it comes to a personalized item recommendation system, It is essential to extract users' preferences and purchasing patterns. 
Assuming that users in the real world form a cluster and there is common favoritism in each cluster, in this work, we introduce \textit{Co-Clustering Wrapper} (CCW). 
We compute co-clusters of users and items with co-clustering algorithms and add CF subnetworks for each cluster to extract the in-group favoritism.
Combining the features from the networks, we obtain rich and unified information about users. 
We experimented real world datasets considering two aspects: Finding the number of groups divided according to in-group preference, and measuring the quantity of improvement of the performance.

\end{abstract}


\section{Introduction}

Collaborative filtering (CF) is a widely used method in recommendation systems  \cite{thorat2015survey}. 
It estimates interactions from given interactions between users and items, and the estimated interactions can be interpreted as estimated values of personal preference of items.
For this reason, CF is used in various industrial areas that need recommendations, such as E-commerce, advertisement, over-the-top media services, news, and social network platforms. 

There are many industrial benchmark datasets for recommendation systems: Movielens  \cite{maxwell_2015}, Amazon (\cite{he2016ups, mcauley2015image}), and Yelp2018 datasets.
Datasets consist of user-item interaction data and additional feature data of users and items. 
The user-item interaction data can be represented by a matrix whose row and column indices indicate users and items, respectively. 
We call this matrix a rating matrix. Since the matrix is so sparse that its entries are almost empty, CF models aim to fill up the empty entries with the given interaction data.
In order to achieve this, various model-based CF methods have been developed in areas of machine learning (\cite{koren2009matrix, polat2005svd, zhou2008large, rendle2012bpr}) and deep learning (\cite{zhou2018deep, guo2017deepfm, barkan2016item2vec, shenbin2020recvae, hidasi2015session, he2017neural, wang2019neural, he2020lightgcn, mao2021ultragcn}).


In the data analysis area, methods of grouping data vectors are called clustering algorithms. 
Although few data types are grouped under experts' knowledge, there is no prior knowledge of most types of data. 
Hence, various clustering algorithms are developed to handle general types of data. 
The categories of clustering methods are vary, such as hierarchical-based \cite{dawyndt2005complete}, fuzzy-based \cite{chang2011fuzzy}, density-based \cite{kumar2016fast}, model-based \cite{van2019kepler}, and so on.
   
In this paper, we propose a novel method called \textit{Co-Clustering Wrapper} (CCW) that wraps a given CF model and make the model capture locality and globality of users and items.
When observing datasets in recommendation systems, we get a motivation of users and items in the dataset.
A type of incidence vectors (rows or columns of a rating matrix) of users or items has globally dense entries, and the other type has globally sparse but locally dense entries. 
We use the \textit{spectral co-clustering} \cite{nie2017learning} in order to make strongly connected clusters consisting of users and items.

The main contributions of our work are followings:
\begin{figure}
    \centering
    \begin{subfigure}[b]{0.98\linewidth}
    \centering
    \includegraphics[width=\linewidth]{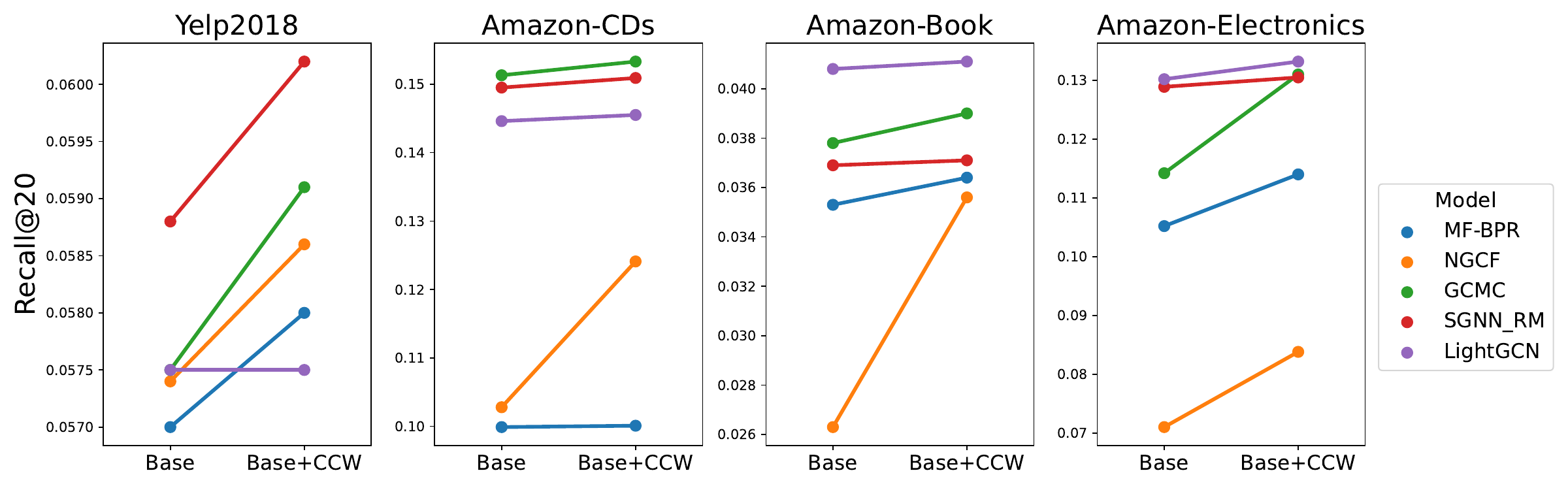}
    \end{subfigure}
    \\
    \begin{subfigure}[b]{0.98\linewidth}
    \centering
    \includegraphics[width=\linewidth]{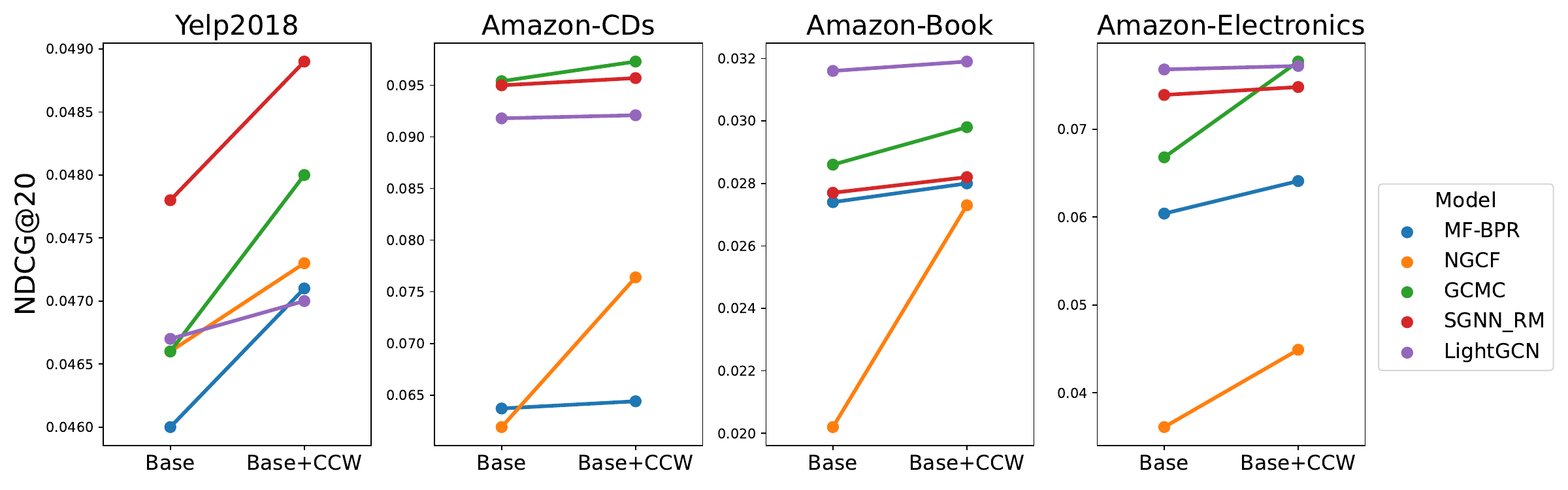}
    \end{subfigure}
    \caption{\label{fig:recall_ndcg_firstpage} \textbf{Comparison of our proposed CCW models and other base models.} Our CCW wrapper reveals and utilizes in-group favoritism in recommendation data, which helps the base model achieve better performance for Recall@20 and NDCG@20 scores.} 
\end{figure}

\begin{enumerate}
    \item We use spectral co-clustering for graph cut of a bipartite graph consisting of users and items. This auxiliary information differs from hand-craft additional features of users and items.
    \item CCW has a university in the sense of independence of choice of CF model. It is because spectral co-clustering can be considered a kind of data preprocessing. Moreover, in most cases, we can observe that CCW improves the performance of CF models.
    \item We suggest a method to determine the number of clusters by using the \textit{variance ratio}.
\end{enumerate}




\section{Related Work}
    \subsection{Graph-Based Recommendation System}
    To tackle graph data and their corresponding tasks (e.g., node classification, relation prediction.), Graph Neural Networks (GNNs) \cite{scarselli2008graph, kipf2016semi} emerged and developed at a rapid pace. In keeping with this, recommendation systems using the user-item bipartite graph structure have been newly studied. 
    
    GC-MC \cite{berg2017graph} first used Graph Convolutional Networks (GCNs) for recommendation systems, regarding matrix completion for recommendation systems as a link prediction on graphs. 
    In addition, information on neighbors based on the rating type is collected and used to configure the new embeddings of the next layer.
    
    NGCF \cite{wang2019neural} exploits embedding refinement, which injects collaborative signal among users and items into their embeddings. Although NGCF obtained better performance than other non-graph models, the high-order convolution process showed limitations in many aspects, including computation efficiency.
    
    Meanwhile, SGNN \cite{yu2021self} tackled stochasticity on GNNs. Stochasticity represents link fluctuations due to external factors, including environment and human factors. By randomly filtering the edges of a given graph, the model aims to construct the GNN, which can handle distributed tasks. 
    
    Recently, studies have continued to simplify the complex and redundant computation of GCNs. SGCN \cite{wu2019simplifying} removed nonlinearities between each layer, making the resulting model linear. Inspired by SGCN, LightGCN \cite{he2020lightgcn} abandoned feature transformation and nonlinearity. The model simply aggregates the connected neighbors by normalization and addition. By reducing the computational burden of the model, it was possible to reduce the adverse effect of the complex propagation sequence of GCNs on the recommendation system. UltraGCN \cite{mao2021ultragcn} skips "infinite" layers of message passing by deriving the infinite powers of message passing.

    There are efforts to set up a standardized benchmark for a graph-based recommendation system. For example, Zhu et al. \cite{zhu2021open} performed open benchmarking for the click-through rate prediction (CTR prediction) task. Zhu et al. \cite{zhu2022bars} also set up the open benchmark for user-item matching tasks. 
    
    \subsection{Co-Clustering}
    Co-clustering, also known as bi-clustering, aims to cluster rows and columns of a matrix simultaneously. The Co-clustering algorithm generates biclusters consisting of a subset of samples and features. By co-clustering, we can fully use the duality information between samples and features in the same bicluster.
    
    Dhillon et al. \cite{dhillon2001co} proposed a spectral algorithm for co-clustering, which is found very efficient. This spectral co-clustering algorithm (SCC) leverages a scaled adjacency matrix's second left and right singular values, minimizing the cut between samples and features.

    Various SCC-based algorithm were developed inspired by \cite{dhillon2001co}, including ensemble-coclustering \cite{affeldt2020ensemble}, CCMOD \cite{ailem2015co}, SCMK \cite{kang2017twin}, ONMTF \cite{abe2019orthogonal}, and SOBG \cite{nie2017learning}.
    Among various co-clustering methods, we adopt fundamental SCC to our CCW model. However, one may use any other co-clustering methods instead of SCC. 

\begin{figure}
    \centering
    \begin{subfigure}[b]{0.32\linewidth}
    \centering
    \includegraphics[width=\linewidth]{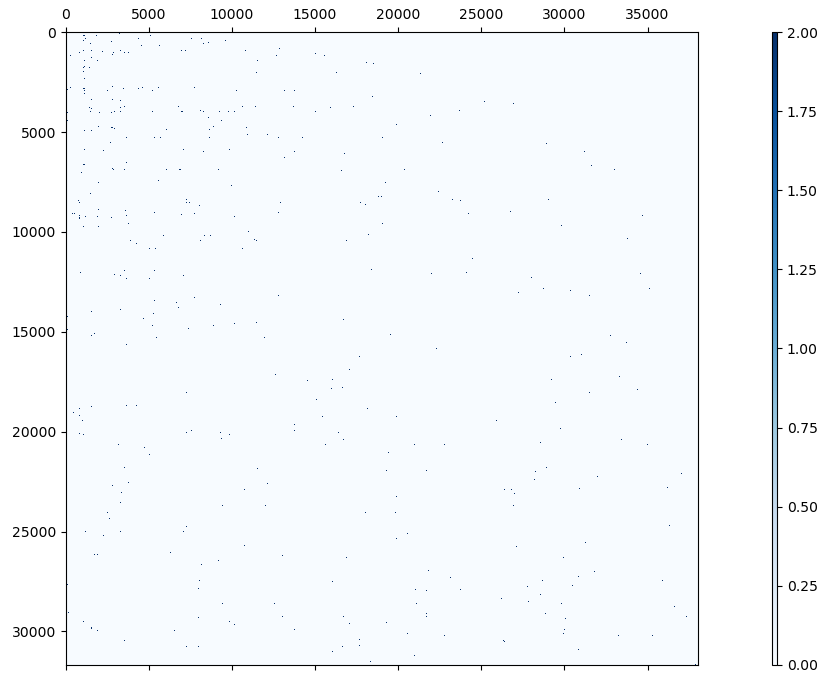}
    \caption{Before SCC}
    \end{subfigure}
    \begin{subfigure}[b]{0.32\linewidth}
    \centering
    \includegraphics[width=\linewidth]{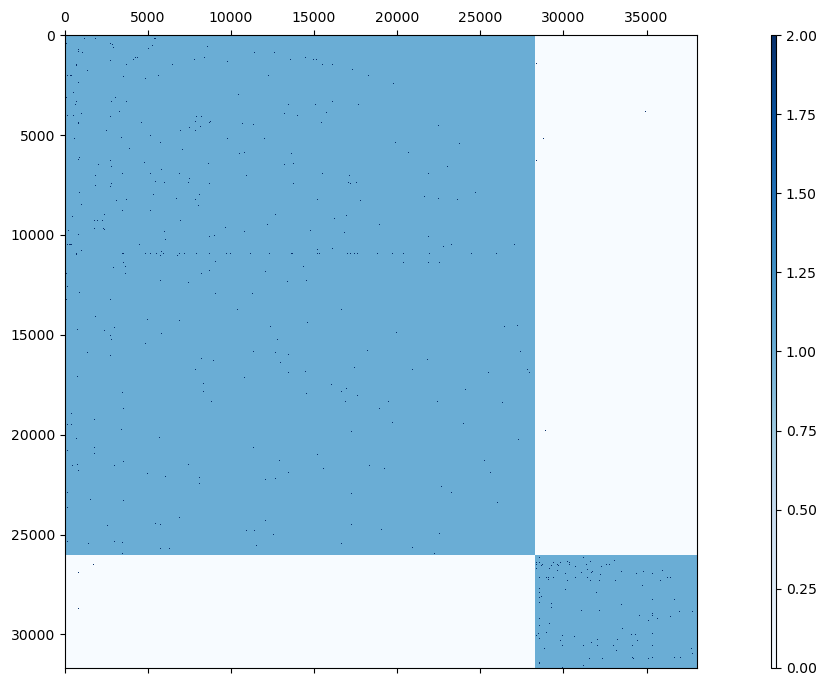}
    \caption{After SCC (k=2)}
    \end{subfigure}
    \begin{subfigure}[b]{0.32\linewidth}
    \centering
    \includegraphics[width=\linewidth]{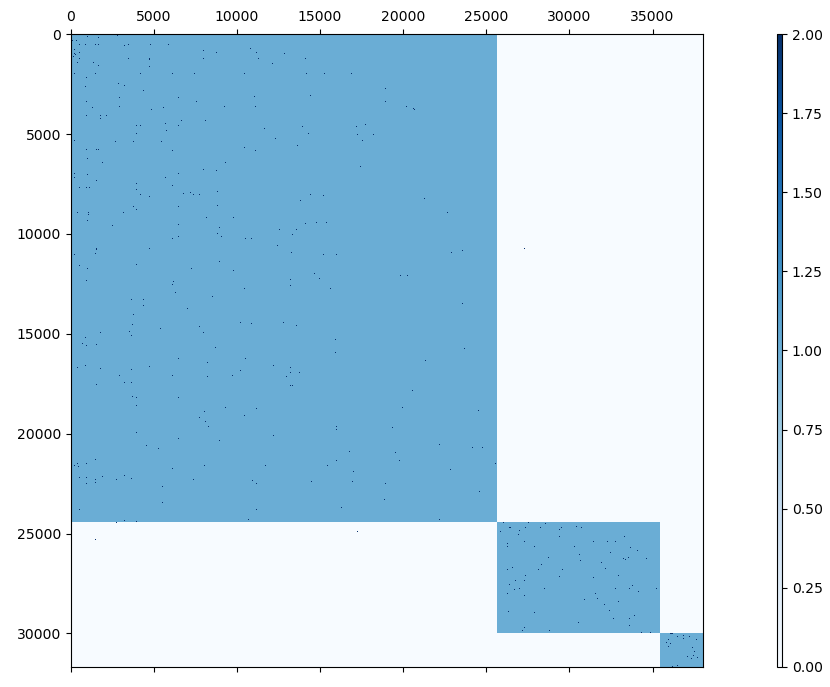}
    \caption{After SCC (k=3)}
    \end{subfigure}
    \\
    \centering
    \begin{subfigure}[b]{0.32\linewidth}
    \centering
    \includegraphics[width=\linewidth]{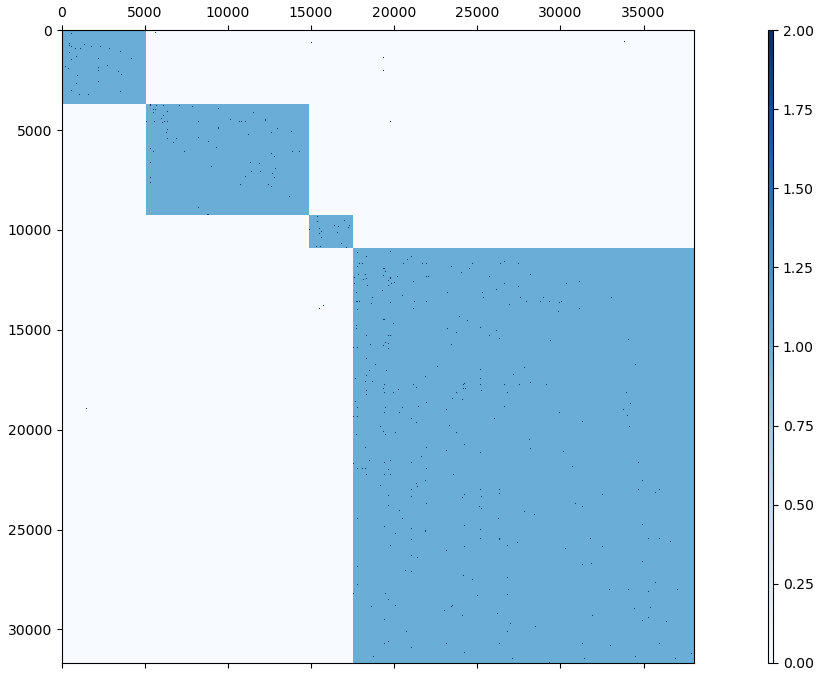}
    \caption{After SCC (k=4)}
    \end{subfigure}
    \begin{subfigure}[b]{0.32\linewidth}
    \centering
    \includegraphics[width=\linewidth]{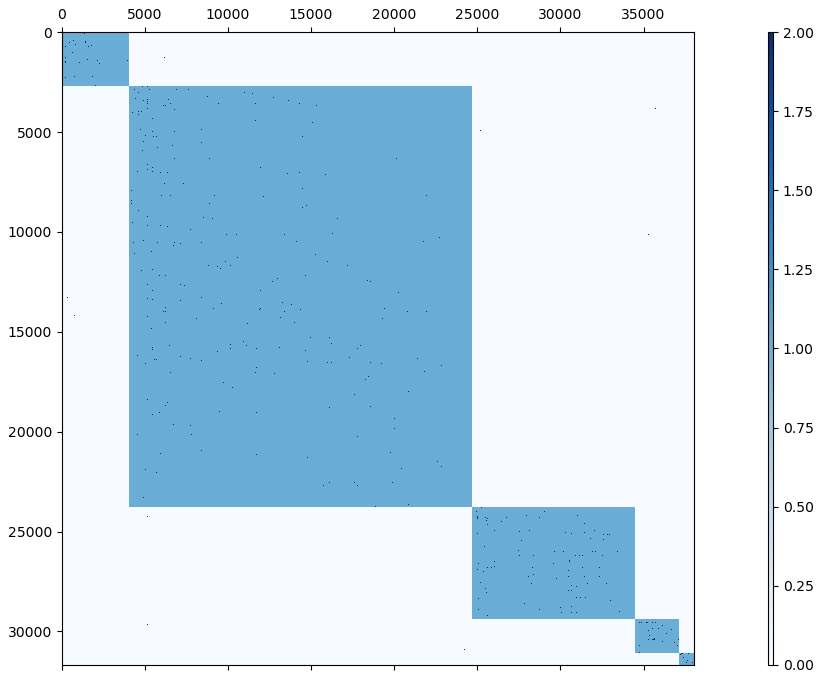}
    \caption{After SCC (k=5)}
    \end{subfigure}
    \begin{subfigure}[b]{0.32\linewidth}
    \centering
    \includegraphics[width=\linewidth]{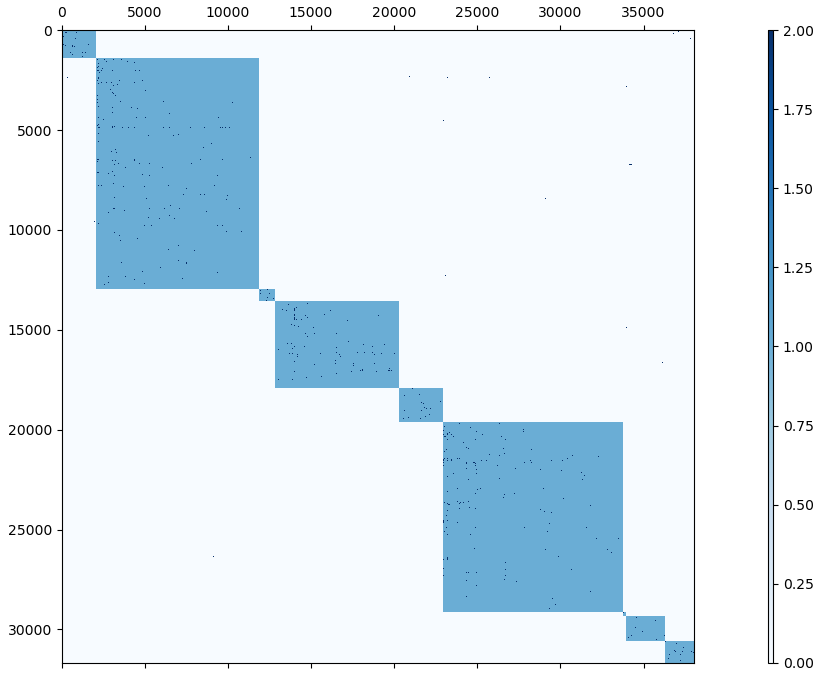}
    \caption{After SCC (k=9)}
    \end{subfigure}
    \caption{
    \label{cocluseter_blocks}\textbf{Co-clustering results of incidence matrix for Yelp2018 Dataset.} For each plot, rows represent users, and columns represent items. (a) shows the plot of the incidence matrix of Yelp2018. From (b) to (f), we show the clustered results according to the value of $K$, which is the number of clusters we want to make. Each blue area in the plot shows the computed co-cluster. 
    }
\end{figure}
    
\section{Methodology}
\begin{figure*}
    \centering
    \includegraphics[width = 0.85\textwidth]{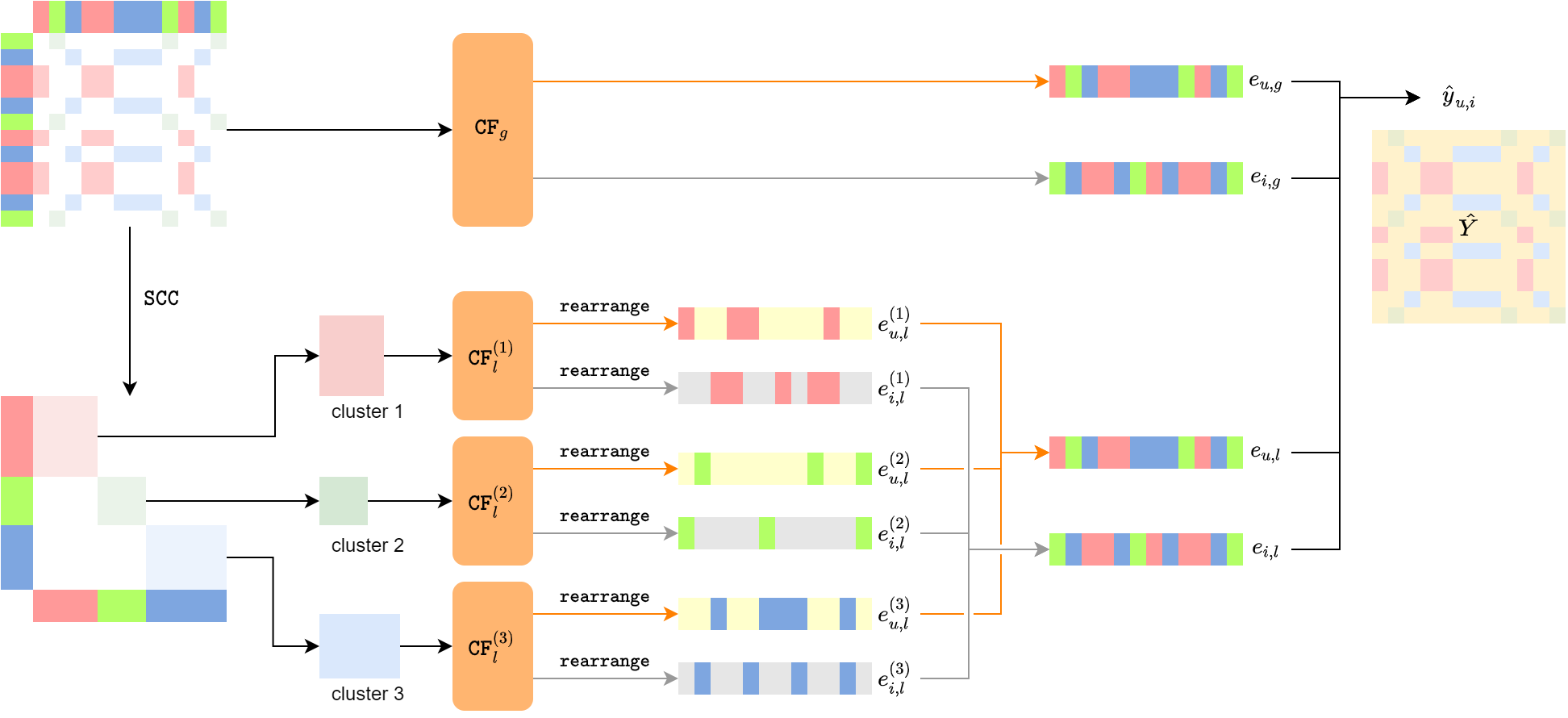}
    \caption{\textbf{Overall flow chart of \textit{Co-Clustering Wrapper}.} First, we apply SCC to the rating matrix, obtaining $k$ co-clusters. (In this figure, we set by $k=3$.) Next, we apply CF models to the entire rating matrix and co-clustered rating matrix in parallel to compute the global and local embeddings. After rearranging the local embeddings to the original order, we aggregate the local and global embeddings, then compute the estimated rating $\hat{Y}$. }
    \label{abs:model}
\end{figure*}

\subsection{Spectral Co-Clustering}
We assume that a large dataset of user-item interactions consists of several clusters which have a strong user-item relationship in each cluster. We can visualize this by comparing incidence matrices of original data and clustered data. 

As shown in Fig. \ref{cocluseter_blocks}, the strong connectivity is shown as a block-diagonal matrix by rearranging indices of rows and columns. 
Clustered user set $u_k$ and item set $i_k$ is defined by
$$    \mathbf{U} = [u_1 \vert u_2 \vert \cdots \vert u_k],$$
    $$\mathbf{I} = [i_1 \vert i_2 \vert \cdots \vert i_k],$$
where $\mathbf{U}$ and $\mathbf{I}$ denote a set of total users and items, respectively. 
\newline
\indent

\subsection{Variance Ratio}
Setting the proper number of clusters $k$ for each dataset is necessary.
If $k$ is too small, a cluster may contain nodes with distinctly different properties.
Also, if $k$ is too large, some over-separated clusters may have the same characteristics. This is inefficient in terms of the purpose of the method.
But computing the number of smallest eigenvalue to find proper $k$ requires tremendous computational cost \cite{dhillon2001co}. 
Therefore, we conducted empirical experiments to find best $k$ using the concept of \textit{variance ratio} \cite{ackerman2009clusterability}.

A variance of a set $X$ is 
\begin{equation}
    \sigma^2(X) = \frac{1}{|X|} \sum_{x \in X} {\lVert x - \text{centroid}(X) \rVert}^2 ,
\end{equation}
where a centroid of a set $X$ is 
\begin{equation}
    \text{centroid}(X) = \frac{1}{|X|} \sum_{x \in X} {x}.
\end{equation}

Let $k$-clustering of $X$ be $C = \{X_1, X_2, ..., X_k\}$, and let $p_i = {|X_i|}/{|X|}$. We compute within-cluster variance $W_C(X)$ and between-cluster variance $B_C(X)$ as follows:
\begin{equation}
    W_C(X) = \sum_{i=1}^{k} p_i \sigma^2(X_i),
\end{equation}
\begin{equation}
    B_C(X) = \sum_{i=1}^{k} p_i {\lVert \text{centroid}(X_i) - \text{centroid}(X) \rVert}^2 .
\end{equation}
Then the variance ratio of clustering $C$ of $X$ is computed as
\begin{equation}
    \operatorname{VR}(C,X) = \frac{B_C(X)}{W_C(X)}.
\end{equation}
High $B_C$ means that clusters are separated from each other, and low $W_C$ implies that the nodes are close to the cluster's centroid. Therefore, a high variance ratio means better clustering quality.

To calculate the variance ratio of the recommendation dataset, we go through the following process.
First, we apply SCC to a given user-item matrix, obtaining user-item clusters.
Then we generate the feature vector of users based on interaction data between all items.
Finally, we calculate the variance ratio of user nodes with clusters and feature vectors computed above.

\begin{figure}[htbp]
    \centerline{\includegraphics[width=0.9\columnwidth]{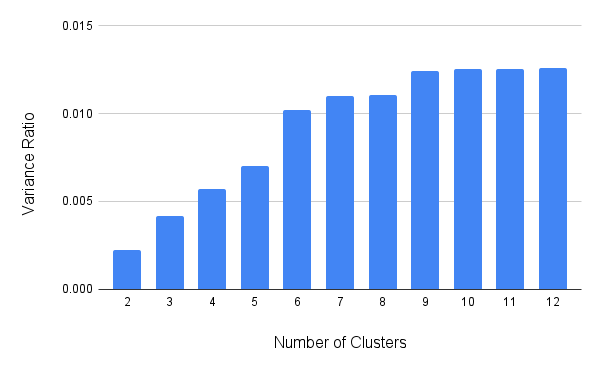}}
    \caption{\textbf{Plot of mean variance ratio of $k$-clustering of Yelp2018.} As $k$ increases, so does the variance ratio. One can discover that the variance ratio hardly changes if $k\geq 9$. Based on the plot, we conclude that $k=9$ is a proper choice for clustering. }
    \label{vr_yelp}
\end{figure}

Fig. \ref{vr_yelp} shows an example of the mean-variance ratios measured by the Yelp2018 dataset for $2 \leq k \leq 12$. For each $k$, we applied spectral co-clustering with $10$ different random seeds and averaged each computed variance ratio.
The variance ratio appears to increase as $k$ increases. 
However, for $k\geq 9$, the variance ratio slightly increases compared to other $k$. 
Considering the efficiency of computing the spectral co-clustering algorithm, We may choose $k=9$ as a clustering number for Yelp2018.
Note that using other random seeds did not significantly change the variance ratio.

\begin{figure*}
    \begin{subfigure}[b]{0.23\linewidth}
    \centering
    \includegraphics[width=\linewidth]{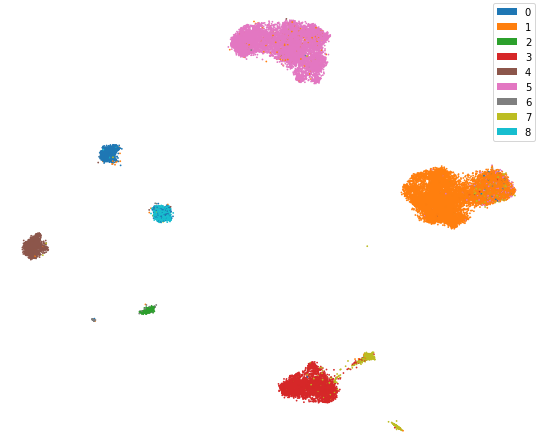}
    \caption{User embeddings via UMAP}
    \end{subfigure} 
    \quad
    \begin{subfigure}[b]{0.23\linewidth}
    \centering
    \includegraphics[width=\linewidth]{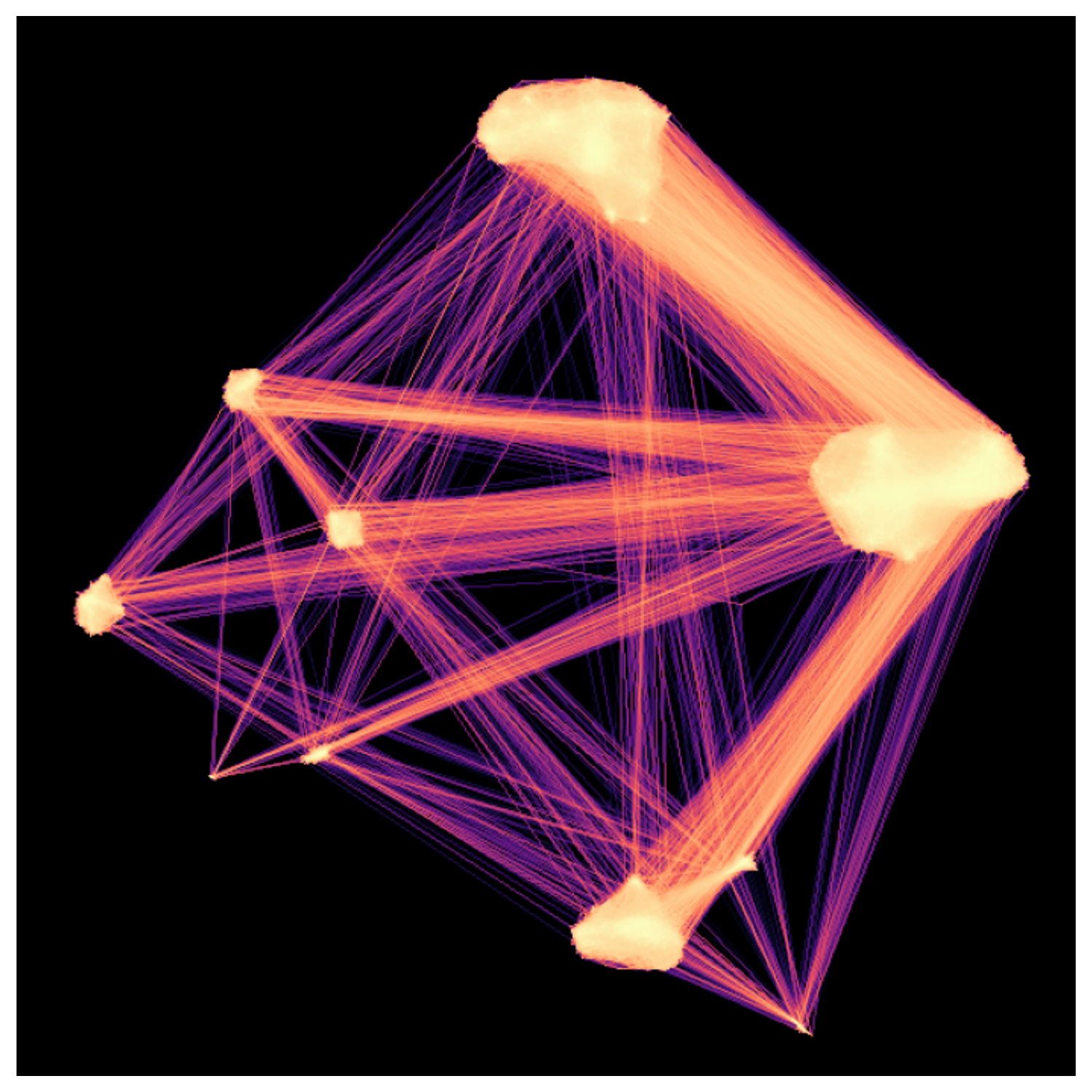}
    \caption{Connectivity plot of Users}
    \end{subfigure}
    \quad
    \begin{subfigure}[b]{0.23\linewidth}
    \centering
    \includegraphics[width=\linewidth]{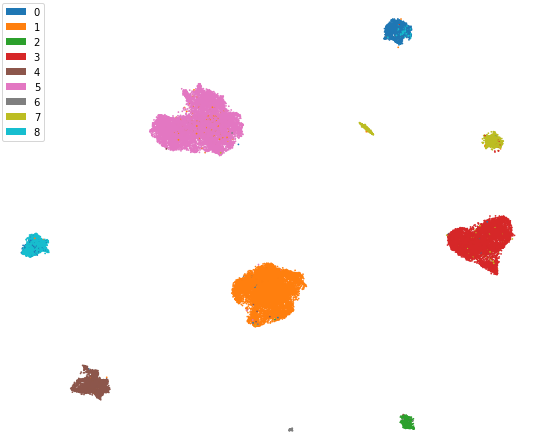}
    \caption{Item embeddings via UMAP}
    \end{subfigure}
    \quad
    \begin{subfigure}[b]{0.23\linewidth}
    \centering
    \includegraphics[width=\linewidth]{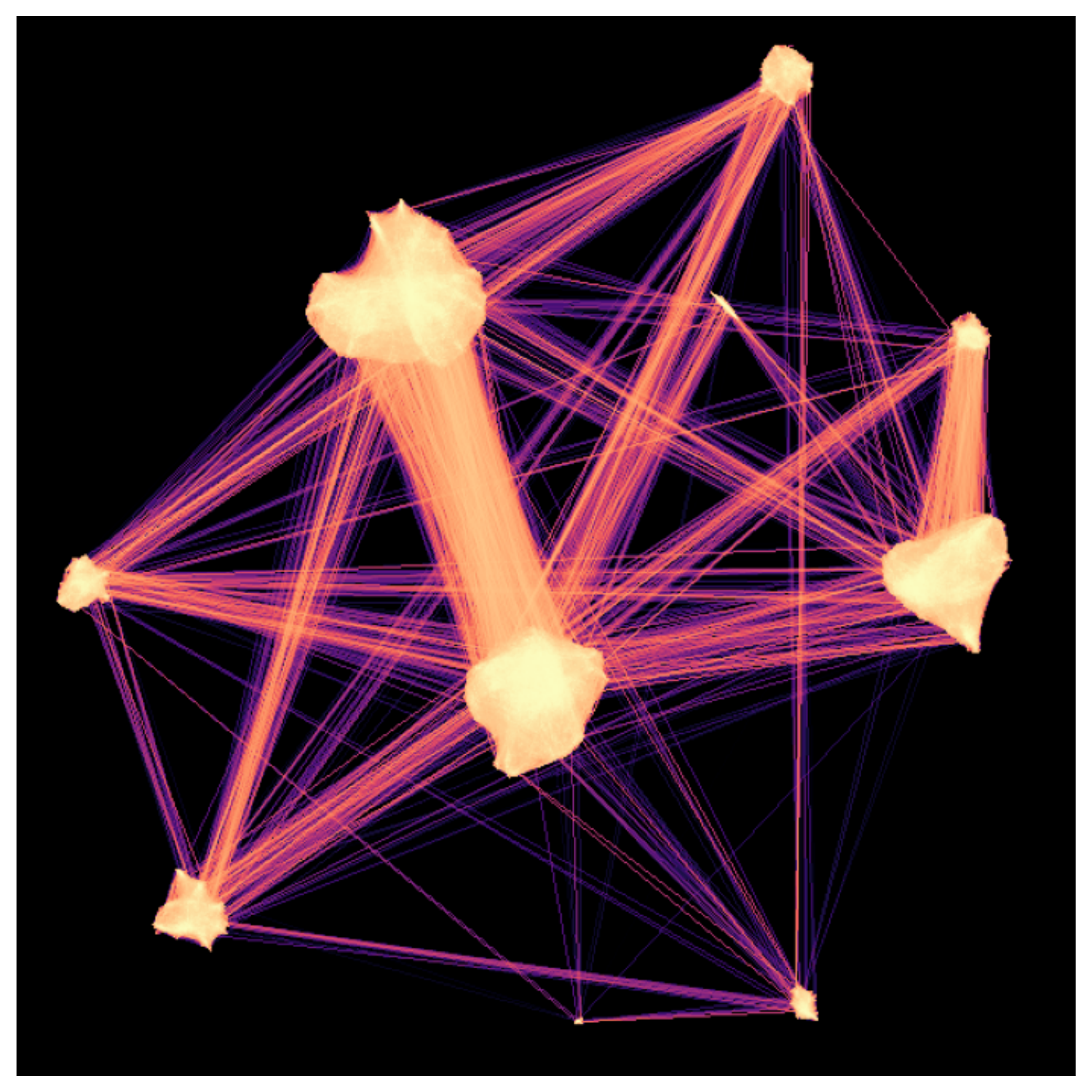}
    \caption{Connectivity plot of Users}
    \end{subfigure} 
    \caption{
    \label{umap_figs}\textbf{UMAP Results for Yelp2018.} The first two figures show the result of applying UMAP to nodes of items of the dataset. The last two figures are the results from the nodes of items. For (a) and (c), points with the same color belong to the same cluster. These clusters are obtained from the result of spectral co-clustering. We find that the colors of the points almost coincide with the computed co-clusters. 
    }
\end{figure*}

In Fig. \ref{umap_figs}, we visualize the UMAP embeddings of the feature vector of users. We use the same feature vector of users in the case of spectral co-clustering. 
The points with the same color are in the same cluster. Therefore, we find that the cluster computed by UMAP almost corresponds to the colors of the points. From this point of view, we conclude that the clustering number $k=9$ is an excellent choice for clustering.

\subsection{Co-Clustering Wrapper}
Our proposed Co-Clustering Wrapper is a method of extending the baseline CF model.
A recommendation dataset which is used in our method can be represented as a bipartite graph $G = (V, E)$. $V$ is a collection of nodes of users and items.
Let nodes of users and items be $V_{\text{item}}$ and $V_{\text{user}}$, respectively. 
$E$ is a collection of edges, denoting interaction between a user and an item. 
First, a cluster $C = \{V_1, ..., V_k \}$ is obtained by the spectral co-clustering method on the entire graph G. Then for each cluster $V_j$ ($j\in \left\{ 1,2,\cdots,k  \right\}$), we generate in-cluster connection $E_j$ by 
\begin{equation}
    E_j = \left\{ (u,v)| (u,v)\in E, u\in V,v\in V\right\}.
\end{equation}
Finally, a subgraph $G_j$ is generated by $G_j = (V_j, E_j)$ for $j\in \left\{ 1,2,\cdots,k  \right\}$.

\subsubsection{Global model and Local models}
There is a baseline CF model which operates on entire graph $G$.
We call this CF model as global model, and denote $CF_g$ to represent global model.
In our CCW algorithm, we additionally use $k$ number of CF models which oprate on each $k$ subgraphs. We call those CF models as local model, and use $CF_l^{(m)}$ to represent $m$-th local model. 
So there are $k+1$ number of models, and they don't share learnable parameters.
By global model, we can get the embedding vector of all nodes, and we call it as global embedding. In the same way, local embedding is obtained by local models.
Since each node belongs to only one cluster, a single global embedding and local embedding can be obtained for each node.
We use $e_{p,g}$ and $e_{p,l}$ to represent global embedding and local embedding of node $p$ respectively.

\subsubsection{Ranking Score between User and Item}
When calculating the ranking score between user $u$ and item $i$, we need to check whether $u$ and $i$ belong to the same cluster. 
If $u$ and $i$ are in different clusters, the local embeddings of $u$ and $i$ are obtained from different local models; therefore, only global embedding is used to calculate ranking scores between them.
So we compute ranking score of item $i$ of user $u$ as follows:
\begin{equation}
\label{eqn:rating score without LIC}
    \hat{y}_{u,i} = \begin{cases} e_{u,g}^{\tp} e_{i,g} + e_{u,l}^{\tp} e_{i,l} & \text{, if } u\sim i\\
    e_{u,g}^{\tp} e_{i,g} & \text{, if } u\not\sim i
    \end{cases},
\end{equation}
where $u\sim i$ denotes that $u$ and $i$ are in the same cluster. 

\subsubsection{Local Importance Coefficient}
If $u$ and $i$ are in the same cluster, we compute each node's final embedding, combining the local and global embedding from the models. 
However, the influence of the local model on nodes in the same cluster may be different. To handle and learn the impact of the local model, we introduce \textit{local importance coefficient} (LIC).
$\text{LIC}$ is a scalar value representing the local model's influence relative to the global model.
We compute the local importance coefficient of node $p \in V$ by simply applying 2-layer MLP to local and global embedding:
\begin{equation}
    \operatorname{LIC}_p = \operatorname{MLP}\left(\left[e_{p,g} \vert e_{p,l}\right]\right)\in \mathbb{R} .
\end{equation}

Finally, the ranking score between user $u$ and item $i$ is computed as follows:
\begin{equation}
\label{eqn:rating score}
    \hat{y}_{u,i} = \begin{cases} e_{u,g}^{\tp} e_{i,g} + (\text{LIC}_u \cdot e_{u,l})^{\tp}(\text{LIC}_i \cdot e_{i,l} ) & \text{, if } u\sim i\\
    e_{u,g}^{\tp} e_{i,g} & \text{, if } u\not\sim i
    \end{cases}.
\end{equation}
Rating score matrix $Y$ is a matrix consisting scores between all users and items.
\begin{equation}
    Y = (\hat{y}_{u,i})_{u, i},
\end{equation}
for $u\in \mathbf{U}$ and $i \in \mathbf{I}$.
We use this ranking scores to top-N recommendation, and also to compute the training loss.

\subsection{Training the Model}
To learn parameters of global model and local models, we optimize BPR loss \cite{rendle2012bpr} as follows:
\begin{equation}
    L_{\text{BPR}} = - \sum_{u=1}^M \sum_{i \in \mathcal{N}_{u}} \sum_{j \notin \mathcal{N}_u} \ln{ \sigma (\hat{y}_{u,i} - \hat{y}_{u,j} ) }+ \lambda \Vert \mathsf{W} \Vert ^2 ,
\end{equation}
where $\mathsf{W}$ is a set of entire learnable parameters of models and $\mathcal{N}_u$ is a set of items which interacted with user $u$.
Through this optimization process, it is possible to learn global and local models at the same time.

\begin{frame}

\begin{algorithm}[H]
\begin{algorithmic}[1]
\Require Bipartite graph of interaction between users and items $G = (V, E)$, the number of cluster $N$
\Ensure Rating score matrix between users and items $Y$
\State Generate subgraphs $\{G_1, ..., G_N\}$ with SCC
\State Compute global embedding of nodes in $G$
\For{$k = 1$ to $N$}
\State Compute local embeddings of nodes in $G_k$
\EndFor
\State Compute LIC of nodes in $G$
\State Compute $Y = (y_{u,i})$ by equation (\ref{eqn:rating score})
\end{algorithmic}
\caption{Overall Algorithm of CCW. }
\label{CCW alg}
\end{algorithm}
\end{frame}

\begin{table}[h]
\centering
\caption{\label{tab:datasetstats}\textbf{Statistics of the datasets.}}
\begin{tabular}{@{}l|r|r|r|r@{}}
\toprule
\multicolumn{1}{c|}{Dataset} & \multicolumn{1}{c|}{\#Users} & \multicolumn{1}{c|}{\#Items} & \multicolumn{1}{c|}{\#Interactions} & \multicolumn{1}{c}{Density} \\ \midrule
Amazon-Book                  & 52,643                       & 91,559                       & 2,984,108                           & 0.00062                     \\
Amazon-CDs                   & 43,169                       & 35,648                       & 777,426                             & 0.00051                     \\
Amazon-Electronics           & 1,435                        & 1,522                        & 35,931                              & 0.01645                     \\
Yelp2018                     & 31,668                       & 38,048                       & 1,561,406                           & 0.00130                     \\ \bottomrule
\end{tabular}
\end{table}
\begin{table*}
\centering
\caption{\label{tab:experiment} \textbf{Experiment Results comparing the base model and CCW model.} We experimented with base and CCW models with five other baseline models and four other datasets. For each dataset, we compute Recall@20 and NDCG@20 as evaluation metrics. The best scores are written in bold for each dataset and evaluation metric. For each baseline models, the evaluation metrics increases after combining CCW. } %
\begin{tabular*}{\textwidth}{|cc|@{\extracolsep{\fill}}cc|cc|cc|cc|}
\hline
\multicolumn{2}{|c|}{Dataset}                               & \multicolumn{2}{c|}{Yelp2018}                          & \multicolumn{2}{c|}{Amazon-CDs}                         & \multicolumn{2}{c|}{Amazon-Book}                       & \multicolumn{2}{c|}{Amazon-Electronics}                       \\ \hline
\multicolumn{2}{|c|}{Model}                                 & \multicolumn{1}{c|}{Recall@20}       & NDCG@20         & \multicolumn{1}{c|}{Recall@20}       & NDCG@20         & \multicolumn{1}{c|}{Recall@20}       & NDCG@20         & \multicolumn{1}{c|}{Recall@20}       & NDCG@20         \\ \hline
\multicolumn{1}{|c|}{\multirow{2}{*}{MF-BPR}}   & Base      & \multicolumn{1}{c|}{0.0570}          & 0.0460          & \multicolumn{1}{c|}{0.0999}          & 0.0637          & \multicolumn{1}{c|}{0.0353}          & 0.0274          & \multicolumn{1}{c|}{0.1052}          & 0.0604          \\ \cline{2-10} 
\multicolumn{1}{|c|}{}                          & CCW & \multicolumn{1}{c|}{\textbf{0.0580}}          & \textbf{0.0471}          & \multicolumn{1}{c|}{\textbf{0.1001}}          & \textbf{0.0644}          & \multicolumn{1}{c|}{\textbf{0.0364}}          & \textbf{0.0280}          & \multicolumn{1}{c|}{\textbf{0.1140}}          & \textbf{0.0641}          \\ \hline
\multicolumn{1}{|c|}{\multirow{2}{*}{NGCF}}     & Base      & \multicolumn{1}{c|}{0.0574}          & 0.0466          & \multicolumn{1}{c|}{0.1028}          & 0.0619          & \multicolumn{1}{c|}{0.0263}          & 0.0202          & \multicolumn{1}{c|}{0.0710}          & 0.0361          \\ \cline{2-10} 
\multicolumn{1}{|c|}{}                          & CCW & \multicolumn{1}{c|}{\textbf{0.0586}}          & \textbf{0.0473}          & \multicolumn{1}{c|}{\textbf{0.1241}}          & \textbf{0.0764}          & \multicolumn{1}{c|}{\textbf{0.0356}}          & \textbf{0.0273}          & \multicolumn{1}{c|}{\textbf{0.0838}}          & \textbf{0.0449}          \\ \hline
\multicolumn{1}{|c|}{\multirow{2}{*}{GCMC}}     & Base      & \multicolumn{1}{c|}{0.0575}          & 0.0466          & \multicolumn{1}{c|}{0.1513}          & 0.0954          & \multicolumn{1}{c|}{0.0378}          & 0.0286          & \multicolumn{1}{c|}{0.1142}          & 0.0668          \\ \cline{2-10} 
\multicolumn{1}{|c|}{}                          & CCW & \multicolumn{1}{c|}{\textbf{0.0591}}          & \textbf{0.0480}          & \multicolumn{1}{c|}{\textcolor{red}{\textbf{0.1533}}} & \textcolor{red}{\textbf{0.0973}} & \multicolumn{1}{c|}{\textbf{0.0390}}          & \textbf{0.0298}          & \multicolumn{1}{c|}{\textbf{0.1310}}          & \textcolor{red}{\textbf{0.0777}}          \\ \hline

\multicolumn{1}{|c|}{\multirow{2}{*}{SGNN\_RM}} & Base      & \multicolumn{1}{c|}{0.0588}          & 0.0478          & \multicolumn{1}{c|}{0.1495}          & 0.0950          & \multicolumn{1}{c|}{0.0369}          & 0.0277          & \multicolumn{1}{c|}{0.1289}          & 0.0739          \\ \cline{2-10} 
\multicolumn{1}{|c|}{}                          & CCW & \multicolumn{1}{c|}{\textcolor{red}{\textbf{0.0602}}} & \textcolor{red}{\textbf{0.0489}} & \multicolumn{1}{c|}{\textbf{0.1509}}          & \textbf{0.0957}          & \multicolumn{1}{c|}{\textbf{0.0371}}          & \textbf{0.0282}          & \multicolumn{1}{c|}{\textbf{0.1305}} & \textbf{0.0748}          \\ \hline

\multicolumn{1}{|c|}{\multirow{2}{*}{LightGCN}} & Base      & \multicolumn{1}{c|}{0.0575}          & 0.0467          & \multicolumn{1}{c|}{0.1446}          & 0.0918          & \multicolumn{1}{c|}{0.0408}          & 0.0316          & \multicolumn{1}{c|}{0.1302}          & 0.0768          \\ \cline{2-10} 
\multicolumn{1}{|c|}{}                          & CCW & \multicolumn{1}{c|}{\textbf{0.0575}}          & \textbf{0.0470}          & 
\multicolumn{1}{c|}{\textbf{0.1455}}          & \textbf{0.0921}          & \multicolumn{1}{c|}{\textcolor{red}{\textbf{0.0411}}}          & \textcolor{red}{\textbf{0.0319}}          & \multicolumn{1}{c|}{\textcolor{red}{\textbf{0.1332}}} & \textbf{0.0772}          \\ \hline

\end{tabular*}
\end{table*}

\begin{figure*}
    \centering
    \begin{subfigure}[b]{0.23\linewidth}
    \centering
    \includegraphics[width=\linewidth]{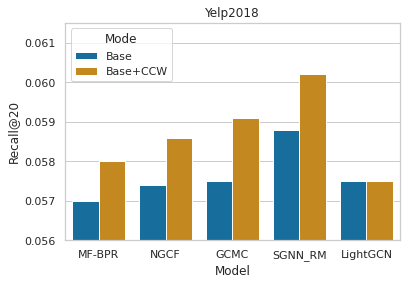}
    \caption{Yelp2018 Recall@20}
    \end{subfigure}
    \quad
    \begin{subfigure}[b]{0.23\linewidth}
    \centering
    \includegraphics[width=\linewidth]{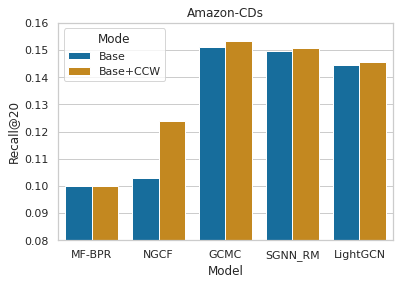}
    \caption{Amazon-CDs Recall@20}
    \end{subfigure}
    \quad
    \begin{subfigure}[b]{0.23\linewidth}
    \centering
    \includegraphics[width=\linewidth]{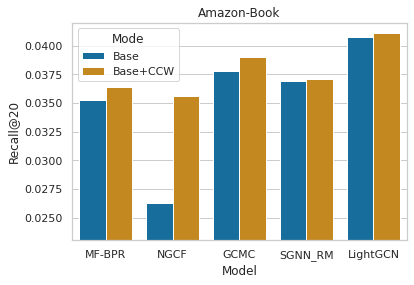}
    \caption{Amazon-Book Recall@20}
    \end{subfigure}
    \quad
    \begin{subfigure}[b]{0.23\linewidth}
    \centering
    \includegraphics[width=\linewidth]{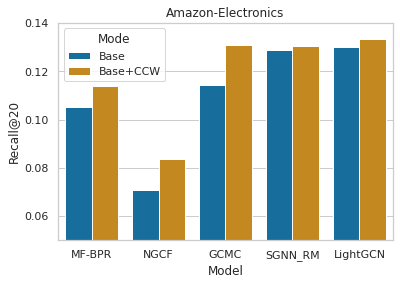}
    \caption{Amazon-Electronics Recall@20}
    \end{subfigure} 
    \\
    \begin{subfigure}[b]{0.23\linewidth}
    \centering
    \includegraphics[width=\linewidth]{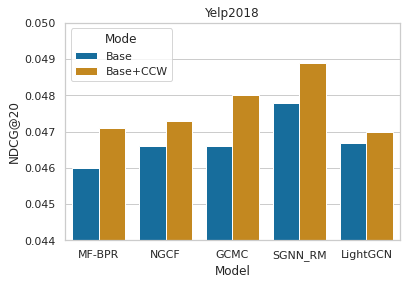}
    \caption{Yelp2018 NDCG@20}
    \end{subfigure} 
    \quad
    \begin{subfigure}[b]{0.23\linewidth}
    \centering
    \includegraphics[width=\linewidth]{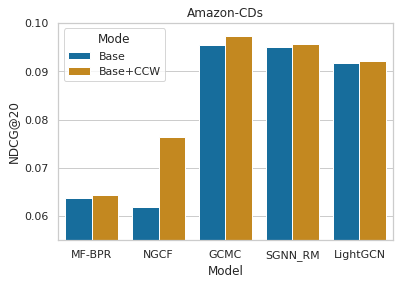}
    \caption{Amazon-CDs NDCG@20}
    \end{subfigure}
    \quad
    \begin{subfigure}[b]{0.23\linewidth}
    \centering
    \includegraphics[width=\linewidth]{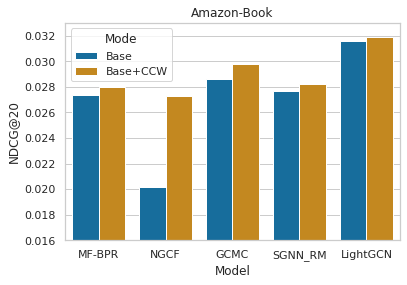}
    \caption{Amazon-Book NDCG@20}
    \end{subfigure}
    \quad
    \begin{subfigure}[b]{0.23\linewidth}
    \centering
    \includegraphics[width=\linewidth]{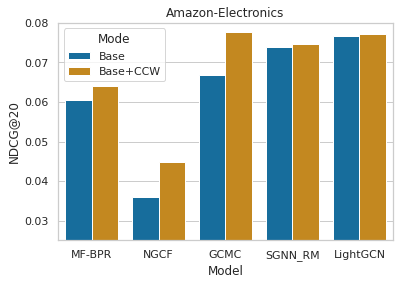}
    \caption{Amazon-Electronics NDCG@20}
    \end{subfigure} 
    \caption{
    \label{experiments_figs} \textbf{Bar graphs of evaluation metrics for each datasets.} In each column, the plots are the results from four other datasets. For each bar graph, the scores are plotted according to what baseline model we used. The results show that the CCW model achieved better scores than the base model.
    }
    \label{fig:bar}
\end{figure*}

\section{Experimental Result}
    \subsection{Dataset and Evaluation Protocol}
    We use four public datasets, including Amazon-Book \cite{he2016ups}, Amazon-CDs, Amazon-Electronics, and Yelp2018, to validate the effectiveness of our model. In addition to our work, other recent graph-based recommendation system models have frequently used these data. For fair model evaluation, data that had already been split for training and test was used. As shown in the Table \ref{tab:datasetstats}, one can see the characteristics of our experimented data having a variety of scales and densities.
    
    For the evaluation protocol, Recall@$K$ and NDCG@$K$  \cite{jarvelin2002cumulated,he2015trirank} were used as evaluation metrics, one for classification and the other for ranking-based metrics respectively. Many other previous papers, \cite{wang2019neural, he2020lightgcn, mao2021ultragcn}, evaluated their models as Recall@20 and NDCG@20, so $K$ was set to 20 as such for fair comparison.
    \subsection{Implementation Details}
    In these experiments, we used various collaborative filtering methods as baseline models. MF-BPR \cite{koren2009matrix}, NGCF \cite{wang2019neural}, GC-MC \cite{berg2017graph}, SGNN \cite{yu2021self}, and LightGCN \cite{he2020lightgcn} were adopted to validate our proposed approach. All the hyperparameters such as initial learning rate, batch size, and regularizer coefficient follow those of the backbone network. Same base model is utilized for local and global collaborative filtering in each experiment. 
    The Adam \cite{kingma2014adam} optimizer was used with $\beta_{1} = 0.9$ and $\beta_{2}=0.999$ as weight decay. 
    \subsection{Result Analysis}
    The experimental results are shown in Fig. \ref{experiments_figs} and Table \ref{tab:experiment}. We adopt \textit{CCW} to four different datasets and five different CF models, including MF-BPR, NGCF, GCMC, SGNN, and LightGCN. Also, Fig \ref{fig:bar} shows the increment bar plot to verify the degree of improvement. Although the performances of the collaborative filtering depend on datasets and CF models, the proposed CCW can enhance the recommendation ability for all cases.
    
    \subsection{Ablation Study}
    \begin{table}[]
    \centering
    \caption{ \label{tab:ablation}\textbf{Ablation Study}}
    \begin{tabular}{|l|l|l|}
        
        \hline
        \textbf{Amazon-Electronics, k=8}      & \textbf{Recall@20} & \textbf{NDCG@20} \\ \hline
        NGCF                         & 0.0849   & 0.0446 \\ \hline
        NGCF + CCW (1:1)             & 0.0666   & 0.0337 \\ \hline
        NGCF + CCW + LIC             & 0.0892   & 0.0472 \\ \hline
  
        \end{tabular}
    \end{table}
    As the proposed method includes LIC in the global-local aggregation part, it is necessary to verify its effect. Therefore, we conducted an ablation study with extensive experiments excluding LIC in the proposed method. Ablation experiments are implemented for the Amazon-Electronics dataset with NGCF, and the number of clusters is set as 8. 
    Experiments are conducted 10 times for each case. 
    \newline
    \indent 
    In this study, excluding LIC corresponds to calculate rating score by equation \ref{eqn:rating score without LIC}, regarding the importance of global and local embedding are equal.
    \newline
    \indent 
    As shown in Table \ref{tab:ablation}, our proposed method can significantly increase the performance. However, the result shows that CCW works well only used with the LIC in global-local aggregation. Furthermore, the performance decreased when the importance of global embedding and local embedding were equal. 
    
\section{Conclusion and Future Work}

In this study, we proposed a recommendation system method using Spectral Co-Clustering. This method co-clusters the given bipartite graphs of interactions between users and items to classify (user, item) pairs into several clusters. Based on the localized bipartite graph of each divided cluster, it was possible to recommend the local flavor of each user rather than using only the default interactions. In the experiment of this paper, we confirmed that this method could further improve the existing graph-based recommendation system model.

However, co-clustering was used as a motivation for a personalized recommendation. The method used in this study is spectral co-clustering, based on spectral analysis of the user-item interaction bipartite graph. Better recommendation results can be expected if a learnable neural network-based co-clustering method is used. 

Additionally, there are studies, including GraphWorld \cite{palowitch2022graphworld}, to generate synthetic graph data without the bias of the data. The study aims to create graphs with various characteristics to evaluate and analyze the GNN model. If we investigate the statistical characteristics of the real-world datasets, we can construct synthetic recommendation datasets with different attributes such as clusterability via these tools. As a result, researchers can develop more abundant model analyses and construct robust models with the generated datasets.

\newpage

\bibliography{IEEEabrv, mybibfile}
\bibliographystyle{IEEEtran}



\end{document}